\documentclass{svproc}

\usepackage{url}

\usepackage{cite}
\usepackage{graphicx}
\usepackage{picinpar}
\usepackage{amsmath}
\usepackage{stfloats}
\usepackage{flushend}
\usepackage[latin1]{inputenc}
\usepackage{colortbl}
\usepackage{soul}
\usepackage{multirow}
\usepackage{pifont}
\usepackage{color}
\usepackage{alltt}
\usepackage{enumerate}
\usepackage{siunitx}
\usepackage{pbox}
\usepackage{amsmath,amssymb,amsfonts}
\usepackage{amsthm}
\usepackage{algorithmic}
\usepackage{algorithm}
\usepackage{graphicx}
\usepackage{graphbox}
\usepackage{textcomp}
\usepackage{xcolor}
\usepackage{colortbl}
\usepackage[export]{adjustbox}
\usepackage{color}
\usepackage{tikz}
\def\BibTeX{{\rm B\kern-.05em{\sc i\kern-.025em b}\kern-.08em
    T\kern-.1667em\lower.7ex\hbox{E}\kern-.125emX}}

\definecolor{LightGreen}{rgb}{0.8,1,0.8}
\definecolor{LightRed}{rgb}{1,0.8,0.8}
\definecolor{LightYellow}{rgb}{1.0,1.0,0.8}

\usepackage{diagbox}

\makeatletter
\let\NAT@parse\undefined
\makeatother
\usepackage{hyperref}

\newcommand\copyrighttext{%
  \footnotesize \textcopyright This paper has been accepted for publication in the Fifth Iberian Robotics Conference (ROBOT22). Please cite the paper as: E. Sebasti\'{a}n, E. Montijano and C. Sag\"{u}\'{e}s,``Multi-robot Implicit Control of Massive Herds'', Fifth Iberian Robotics Conference (ROBOT22), 2022.}
\newcommand\copyrightnotice{%
\begin{tikzpicture}[remember picture,overlay]
\node[anchor=south,yshift=10pt] at (current page.south) {\fbox{\parbox{\dimexpr\textwidth-\fboxsep-\fboxrule\relax}{\copyrighttext}}};
\end{tikzpicture}%
}

\begin{document}
\mainmatter             

\title{Multi-robot Implicit Control of Massive Herds}

\titlerunning{Multi-robot Implicit Control of Massive Herds}  %

\author{Eduardo Sebasti\'{a}n \and Eduardo Montijano \and Carlos Sag\"{u}\'{e}s}
\authorrunning{Eduardo Sebasti\'{a}n et al.} 

\tocauthor{Eduardo Sebasti\'{a}n, Eduardo Montijano and Carlos Sag\"{u}\'{e}s}

\institute{Universidad de Zaragoza, DIIS-I3A, Spain\\
\email{\{esebastian,emonti,csagues\}@unizar.es}}

\maketitle   

\copyrightnotice

\begin{abstract}
This paper solves the problem of herding countless evaders by means of a few robots. The objective is to steer all the evaders towards a desired tracking reference while avoiding escapes. The problem is very challenging due to the highly complex repulsive evaders' dynamics and the underdetermined states to control. We propose a solution that is based on Implicit Control and a novel dynamic assignment strategy to select the evaders to be directly controlled. The former is a general technique that explicitly computes control inputs even in highly complex input-nonaffine dynamics. The latter is built upon a convex-hull dynamic clustering inspired by the Voronoi tessellation problem. The combination of both allows to choose the best evaders to directly control, while the others are indirectly controlled by exploiting the repulsive interactions among them. Simulations show that massive herds can be herd throughout complex patterns by means of a few herders.
\end{abstract}

\section{Introduction}\label{sec:intro}

Herding in multi-robot systems~\cite{sebastian2022adaptive} refers to the problem of steering groups of entities (\textit{evaders}) towards desired positions or trajectories by means of a team of robots (\textit{herders}). The main particularity of herding is that the evaders are \textit{non-cooperative}, in the sense that their behavior is determined by complex repulsive forces. The repulsion models the natural tendency of evaders to escape from possible threats, where the threat is represented by the robotic herders.

From a control perspective, the problem of herding is very challenging because the dynamics of the evaders are typically described by highly complex input-nonaffine equations, which means that the nonlinearities also affect the control inputs. Unlike input-affine nonlinear systems, where many standard control techniques are well-known (e.g., feedback linearization), there is no standard control technique for input-nonaffine systems. As a consequence, herding solutions are usually particular to the specific dynamics they consider~\cite{zhang2022collecting} or assume linear dynamics to facilitate the control~\cite{Scott_ACC_2013_PHE}. Implicit Control~\cite{sebastian2020multi} is a novel general control technique for input-nonaffine systems, which allows to address the herding for general applications. 
This methodology can also handle uncertainties in the evaders' dynamics~\cite{sebastian2022adaptive}, similarly to other model-free~\cite{Licitra_TRO_2019_Single, Gao_ACCESS_2018_Escorting} or model-based herding solutions~\cite{schwarting2021stochastic}. Despite its flexibility, a current limitation of this methodology is that it requires a number of inputs of the same order of the number of controlled states.

In herding, the number of available herders is considerably smaller than the number of evaders.
Many herding solutions are restricted to a single evader~\cite{Monti_CDC_2013_Entrapment, zhi2021learning, franchi2016decentralized}, deal with the evaders one by one~\cite{Licitra_TRO_2019_Single}, assume cooperation~\cite{tsatsanifos2021modeling,zhi2021learning}, or only deal with a similar number of herders and evaders~\cite{sebastian2020multi,sebastian2022adaptive}. Inspired by surveillance and caging-like solutions~\cite{Auletta_AuRo_2022_Herding,song2021herding}, and to overcome the limitation on the number of controlled variables, we propose a novel dynamic assignment strategy to decide online the best evaders to directly control. The rest of the herd is indirectly controlled, but all the evaders converge to the desired tracking reference while ensuring that none of them escape. Moreover, this is done by assuming that the evaders move as fast as the herders, which increases the difficulty of the problem and differs from most of the aforementioned solutions.

The main contributions of this work are a new dynamic assignment strategy (Section~\ref{sec:proposal}) and its combination with Implicit Control (Section~\ref{sec:implicit}) to solve the herding problem of a countless number of evaders with only a few herders (Section~\ref{sec:problem}). This combination extends the results presented by Sebastian et al.~\cite{sebastian2020multi,sebastian2022adaptive} by overcoming the need of a similar number of herders and evaders to achieve the herding. At each instant, the strategy selects as many evaders as herders to be directly controlled towards the desired tracking trajectory, choosing the worthiest ones in terms of coalition of the herd. The strategy leverages the notion of Voronoi tessellation to compute a set of clusters, where the evaders that form the clusters are the ones that constitute the perimeter of the convex hull of the herd. Different simulated experiments (Section~\ref{sec:results}) validate the success of the proposal, concluding that our solution can herd up to hundreds of complex heterogeneous input-nonaffine evaders with only $3-6$ robots (Section~\ref{sec:conclusions}).

\section{Problem Formulation}\label{sec:problem}

Consider a group of evaders $\mathcal{E} = \{ 1,\hdots,j,\hdots , m \}$ and a team of herders $\mathcal{H} = \{1,\hdots,i,\hdots , n\}$. Evaders and herders are described by their positions, $\mathbf{x}_{j}\in\mathbb{R}^{2}$ and $\mathbf{u}_{i}\in\mathbb{R}^{2}$. The position of herders and evaders is stacked to form the state and input of the system
\begin{equation}
\mathbf{x} = 
\begin{bmatrix}
    \mathbf{x}_{1}^{T} & \dots  & \mathbf{x}_{m}^{T}
\end{bmatrix}^{T} 
\hbox{ , }  
\mathbf{u} = \begin{bmatrix}
    \mathbf{u}_{1}^{T} & \dots  & \mathbf{u}_{n}^{T}
    \end{bmatrix}^{T}.
\label{eq:defs}
\end{equation}
The dynamics of each evader is given by
\begin{equation}
\dot{\mathbf{x}}_j = f_j(\mathbf{x},\mathbf{u}) = {f_j^h(\mathbf{x}_j,\mathbf{u})}+ {f_j^e(\mathbf{x})}. 
\label{eq:base}
\end{equation}
We assume that $f_j(\cdot)$ is continuous and with continuous derivative $\forall j \in \mathcal{E}$. The dynamics of evader $j$ are the result of the repulsion provoked by the herders ($f_j^h(\mathbf{x}_j,\mathbf{u})$) and the other evaders ($f_j^e(\mathbf{x})$). 
An example of such model can be the one defined by Pierson and Schwager~\cite{Pierson_2018_TR_Herding},
\begin{equation}
\dot{\mathbf{x}}_j  = {\theta_j\sum_{i=1}^{n} \frac{\mathbf{d}_{ij}}{||\mathbf{d}_{ij}||^{3}}} + {\gamma_j \sum_{j^{\prime}=1, j \neq j^{\prime}}^{m} \frac{\mathbf{d}_{j^{\prime}j}}{||\mathbf{d}_{j^{\prime}j}||^{2}}},
\label{eq:PiersonBase}
\end{equation}
where $\mathbf{d}_{ij} = \mathbf{x}_{j} - \mathbf{u}_{i}$ is the relative position between evader $j$ and herder $i$, $\mathbf{d}_{j^{\prime}j} = \mathbf{x}_{j} - \mathbf{x}_{j^{\prime}}$ is the relative position between evaders $j$ and $j^{\prime}$, $||\;||$ is the L2 norm, and $\theta_j$ and $\gamma_j$ are positive constants which express the aggressiveness in the repulsion provoked by the herders and the other evaders respectively. 

The repulsion term among evaders enhances the realism in the behavior of the evaders because we explicitly avoid collision among evaders. On the other hand, it hinders the control problem because the evaders are also repelled by other evaders, which promotes interaction forces against the desired coalition pursued by the herders.

The overall system dynamics is
\begin{equation}
\dot{\mathbf{x}} = f(\mathbf{x},\mathbf{u})
\label{eq:initial_system},
\end{equation}
where $f(\mathbf{x},\mathbf{u}) = [f_1^T(\mathbf{x},\mathbf{u}), \hdots, f_m^T(\mathbf{x},\mathbf{u})]^T$. This formulation allows to consider heterogeneous herds, with different herd sizes and motion models. We assume that the maximum velocity of both herders and evaders is $v_{\max} > 0$. Besides, we assume that the herders are initially deployed surrounding the evaders. The first assumption hinders the herding problem. The second assumption is reasonable as long as there already exist algorithms~\cite{song2021herding,sebastian2022adaptive} that achieve it.

Finally, regarding the control objective, the goal is to herd the evaders towards a desired position $\mathbf{x}^{*} \in \mathbb{R}^2$ or tracking reference $\dot{\mathbf{x}}^* \in \{\ \dot{\mathbf{x}}^* \in \mathbb{R}^2 \hbox{ } | \hbox{ } ||\dot{\mathbf{x}}^*|| \leq v_{\max} \}$. We assume that $\dot{\mathbf{x}}^*$ is continuous, with continuous derivatives, and bounded. Given Eq.~\eqref{eq:PiersonBase}, it is clear that $\mathbf{x}_j = \mathbf{x}^{*}$ $\forall j \in \mathcal{E}$ is impossible due to the repulsion among evaders. Therefore, the control objective is as follows.

\begin{definition}\label{def:control}
Given $\mathbf{x}$, $\mathbf{u}$, $\mathbf{x}^{*}$, $\dot{\mathbf{x}}^*$, $f(\mathbf{x},\mathbf{u})$, the control strategy $\pi(\mathbf{x},\mathbf{u})$ solves the herding problem if 
\begin{enumerate}
    \item $\mathbf{x}_c = \frac{1}{m}\sum_j^m \mathbf{x}_j$ (the centroid of the herd) converges to $\mathbf{x}^{*}$,
    \item  $\dot{\mathbf{x}}_c = \frac{1}{m}\sum_j^m \dot{\mathbf{x}}_j$ (the speed of the centroid of the herd) converges to $\dot{\mathbf{x}}^*$, and
    \item  $\alpha_j < \lim_{t\rightarrow \infty} ||\mathbf{x}_j - \mathbf{x}_c|| < \beta_j$  $\forall j \in \mathcal{E}$, where $0 <\alpha_j, \beta_j < \infty$ are lower and upper bounds dependent on $f_j^e(\mathbf{x})$ and $f_j^h(\mathbf{x})$.
\end{enumerate}
\end{definition}
The last point of Definition~\ref{def:control} means that the evaders converge to a position as close as possible to the desired tracking reference, considering the repulsion provoked by the other evaders and the herders. In this regard, it is worth noting that the herding seeks a herd formation similar to a flock~\cite{olfati2006flocking}. The significant difference is that in flocking the agents are naturally cooperative and the controller includes both the attractive and repulsive forces among agents to achieve coalition, while in herding only repulsive forces exist, so the robotic herders must move to find the steady-state formation that achieves the herding according to Definition~\ref{def:control}. The upper bound $\beta_j$ is given by the repulsion provoked by the herders, so the herd achieves compactness in the sense that evaders are as close as possible to the centroid; the lower bound $\alpha_j$ is given by the repulsion provoked by other evaders and models the fact that evaders can not move arbitrarily close to the centroid because they will be repelled by the other evaders.

\section{Brief Introduction to Implicit Control}\label{sec:implicit}

The proposal of this work resorts to the Implicit Control theory~\cite{sebastian2020multi,sebastian2022adaptive}, so in this Section we briefly introduce the main concepts behind Implicit Control and link them with the herding problem.

The intuition of Implicit Control is that, in input-nonaffine systems, the analytical expressions for the input yield to a set of implicit equations, which can not be solved in closed-form. This issue is overcome if an expression for the time derivative of the input is found. To do so, let
\begin{equation}\label{eq:def_h_CDC}
    h(\mathbf{x},\mathbf{u}) = f(\mathbf{x},\mathbf{u}) - f^*(\mathbf{x}) - \dot{\mathbf{x}}^*,
\end{equation}
with $f^*(\mathbf{x})$ such that $\dot{\mathbf{x}}$ is stable and encodes the desired behavior of the herd in terms of settling time, transient shape, etc. We assume that $f^*(\mathbf{x})$ is continuous and with continuous derivatives. Finding the roots of $h(\cdot)$ is equivalent to solving the control problem, but since $h(\cdot)$ is an implicit equation, then the existence of solution must be guaranteed. It is shown~\cite{sebastian2022adaptive} that under the assumption that $n \geq m$ the solution exists. But we seek a herding solution that deals with many more evaders than herders. This is one of the main motivations of the proposal in Section~\ref{sec:proposal}; but now, for clarity, we assume $n \geq m$.

The solution consists in considering $h$ as a dynamic system and designing $\dot{\mathbf{u}}$ such that $h$ converges to $0$. If the input $\mathbf{u}$ achieves $h=0$, then $\dot{\mathbf{x}}$ follows $f^*(\cdot) - \dot{\mathbf{x}}^*$; and since $f^*(\cdot)$ is stable, $\mathbf{x}$ converges to the desired tracking reference. If the desired evolution of function $h(\cdot)$ is defined as 
\begin{equation}\label{eq:def_h_star}
    \frac{d h(\mathbf{x},\mathbf{u})}{d t} = h^*(\mathbf{x},\mathbf{u}),
\end{equation}
with $h^*(\cdot)$ such that $h(\cdot)$ is stable, then the input dynamics
\begin{equation}
\label{eq:u_dynamics}
    \dot{\mathbf{u}} = \mathbf{J}_{\mathbf{u}}^{+}
    \left(h^*(\mathbf{x},\mathbf{u})-\mathbf{J}_{\mathbf{x}} f(\mathbf{x}, \mathbf{u}) \right)
\end{equation}
imposes Eq.~\eqref{eq:def_h_star}, so $h(\cdot)$ converges to $\mathbf{0}$. In Eq.~\eqref{eq:u_dynamics}, $\mathbf{J}_{\mathbf{x}}$ is the Jacobian of $h(\cdot)$ with respect to $\mathbf{x}$,  $\mathbf{J}_{\mathbf{u}}$ is the Jacobian of $h(\cdot)$ with respect to $\mathbf{u}$ and $\;^+$ is the pseudoinverse of a matrix. The use of the pseudoinverse is because $m \neq n$ in general; the existence of the pseudoinverse is guaranteed by the aforementioned assumptions~\cite{sebastian2022adaptive}.  Eqs.~\eqref{eq:initial_system} and~\eqref{eq:u_dynamics} yield to the explicit system 
\begin{equation}
\label{eq:diff_eq_system_u}
\left\{
\begin{aligned}
    \dot{\mathbf{x}} &= f(\mathbf{x}, \mathbf{u})
    \\ 
    \dot{\mathbf{u}} &= \mathbf{J}_{\mathbf{u}}^{+}
    \left(h^*(\mathbf{x},\mathbf{u})-\mathbf{J}_{\mathbf{x}} f(\mathbf{x}, \mathbf{u}) \right)
\end{aligned}
\right. .
\end{equation}
The control problem is then reduced to analyzing the stability of the system in~\eqref{eq:diff_eq_system_u}. In particular, in this paper it is considered that
\begin{equation}
\label{Eq:f_star_real}
f^*(\mathbf{x})=-\mathbf{F}\mathbf{x}  \kern 0.5cm \hbox{ and } \kern 0.5cm h^*(\mathbf{x},\mathbf{u}) =
    - \mathbf{H} h(\mathbf{x},\mathbf{u}).
\end{equation}
$\mathbf{F}, \mathbf{H}$ are two matrices chosen such that 
$\begin{pmatrix}
    -\mathbf{F} & 0.5 \mathbf{I}
    \\
    0.5 \mathbf{I} & -\mathbf{H}
    \end{pmatrix}$
is negative definite. Under these conditions, the system in~\eqref{eq:diff_eq_system_u} is globally asymptotically stable~\cite{sebastian2022adaptive}.

It is clear that Implicit Control facilitates the design of controllers for general input-nonaffine systems. This opens the possibility of controlling heterogeneous herds regardless of the complexity of the evaders' dynamics. Nonetheless, the issue with the number of evaders remains, because at this point a similar number of herders and evaders is needed. In the next Section we design a novel strategy to borrow the concepts of Implicit Control and herd countless evaders.

\section{Proposed Solution}\label{sec:proposal}

At each instant, herders under Implicit Control can steer as many evaders as herders. Nevertheless, it is not required, a priori, that these evaders are the same for all time. Our proposal exploits this fact in order to select, at each instant, the most convenient evaders to be directly controlled. Therefore, the constraint in the number of controlled evaders is always preserved, whereas the other evaders are steered indirectly. The proposal is materialized in a dynamic assignment strategy that we describe in what follows.

\begin{figure}[!ht]
    \centering
    \begin{tabular}{cccc}
        (a) 
        &
        (b)
        \\
        \includegraphics[width=0.38\textwidth,height=0.38\textwidth]{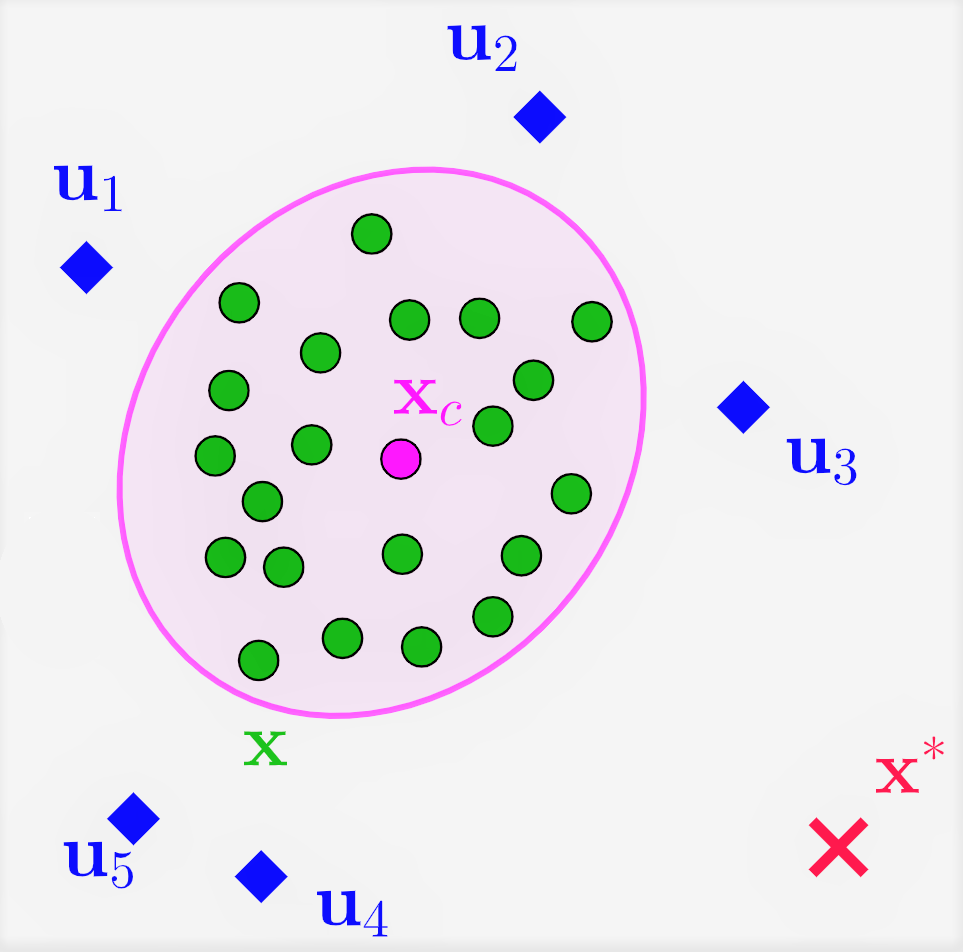}
         &  
         \includegraphics[width=0.38\textwidth,height=0.38\textwidth]{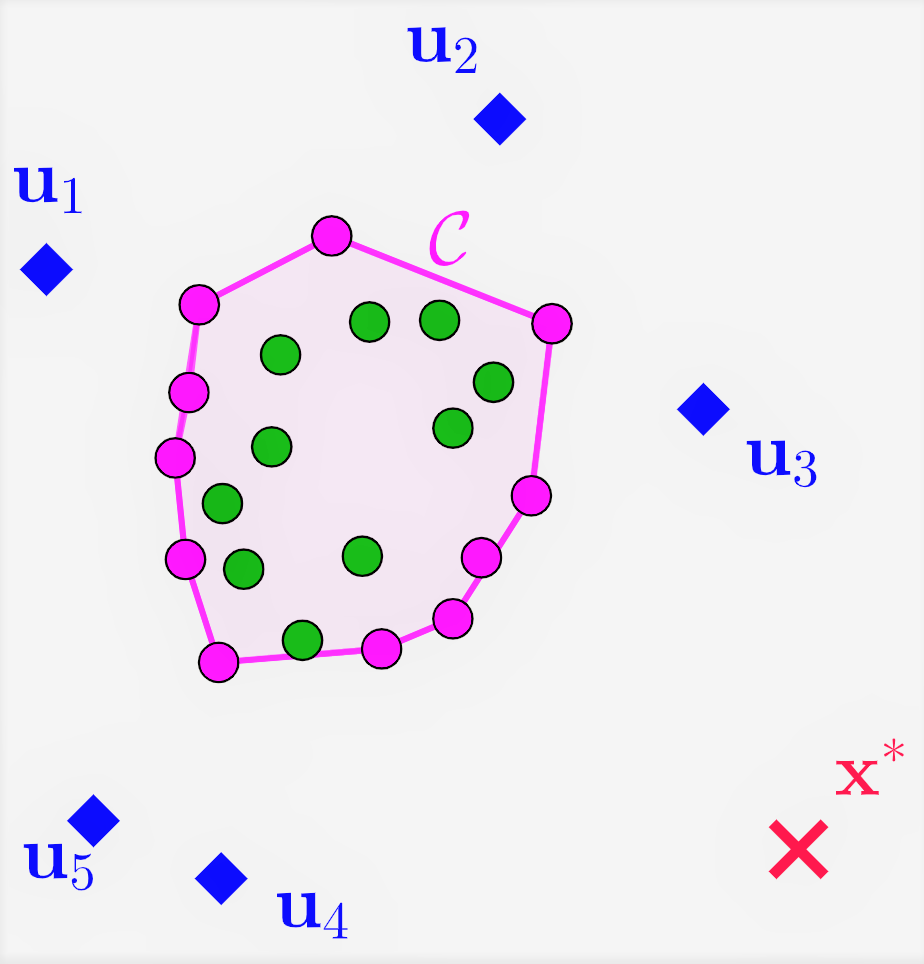}
        \\
        (c)
        &
        (d)
        \\
         \includegraphics[width=0.38\textwidth,height=0.38\textwidth]{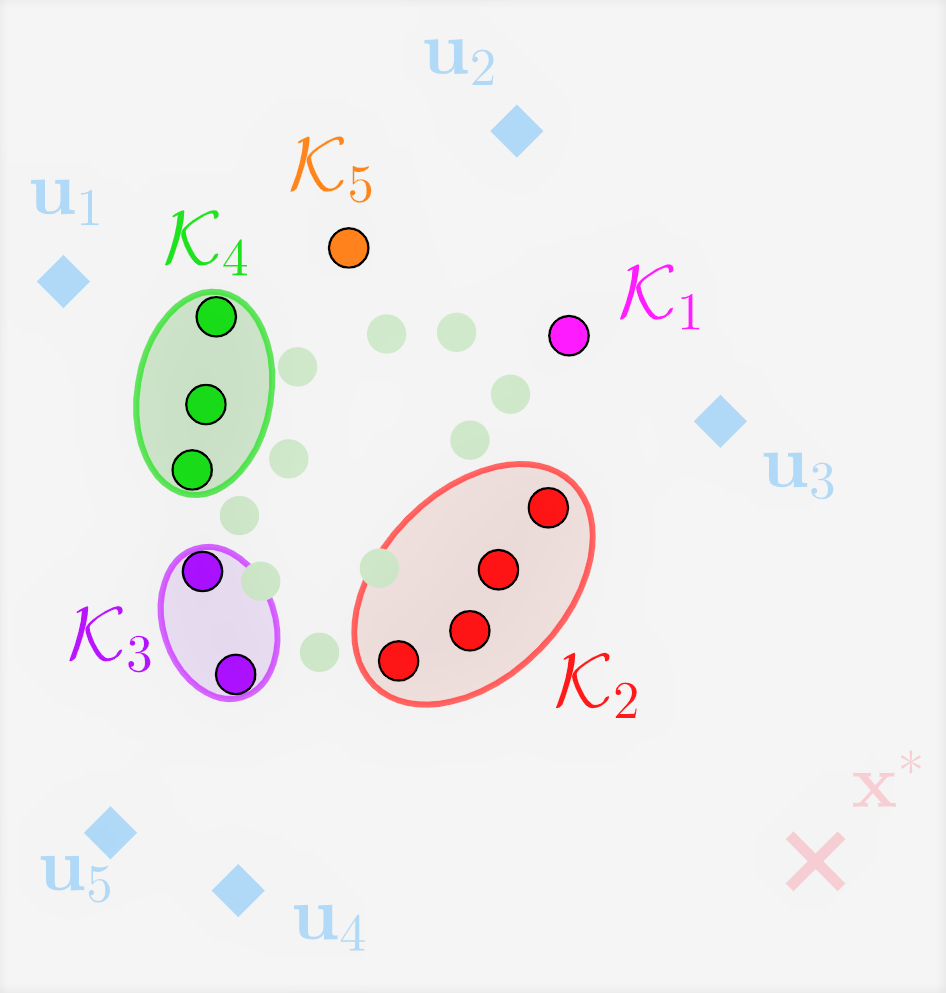}
         &  
         \includegraphics[width=0.38\textwidth,height=0.38\textwidth]{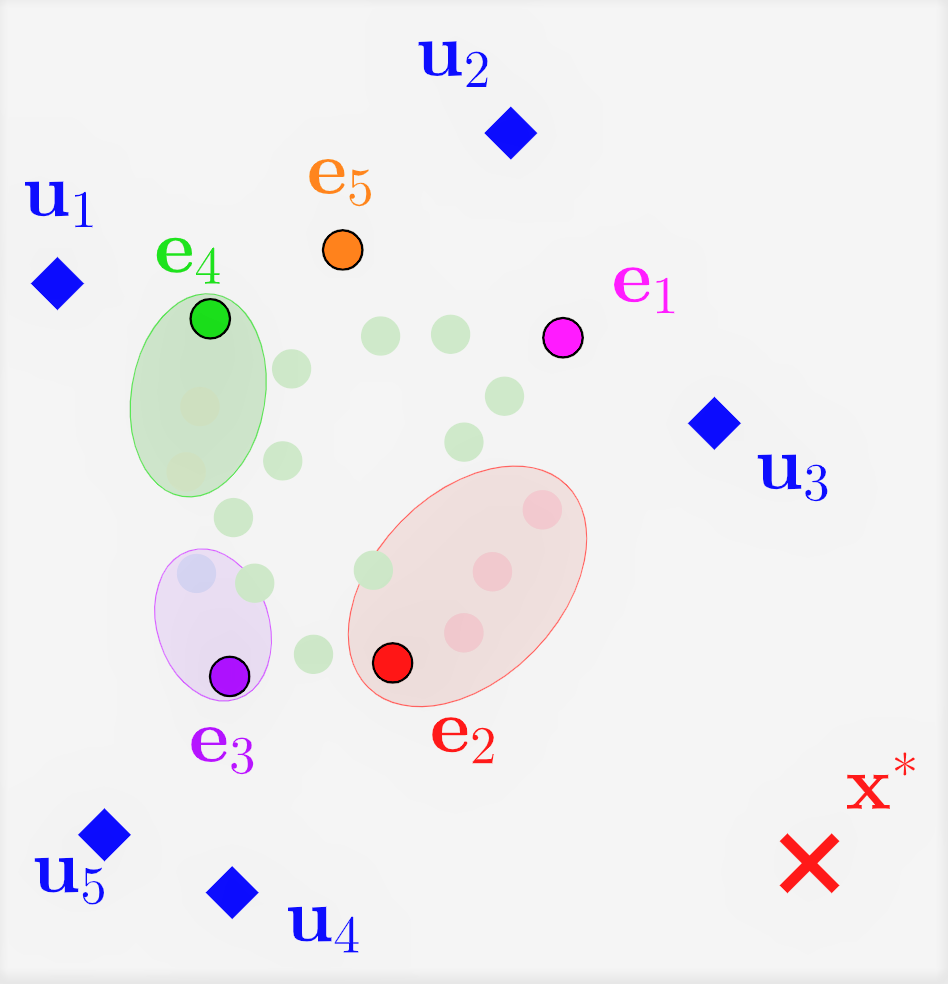}
    \end{tabular}
	\caption{Visualization of the dynamic assignment strategy: (a) configuration at instant $t$, (b) convex hull of the herd, (c) computed clusters, (d) final assignment.}
	\label{fig:assigment}
\end{figure}

The first step is to compute the convex hull of the herd (Fig.~\ref{fig:assigment}a). Let $\mathcal{X} = \{\mathbf{x}_j \}_0^m$ be the set with all the evaders' positions. The convex hull of $\mathcal{X}$ is 
\begin{equation}\label{eq:convex_hull}
    \mathcal{CH} = \left\{ \sum_{j=0}^m \lambda_j\mathbf{x}_j, \: \lambda_j \geq 0 \hbox{ }\forall j \hbox{ and } \sum_{j=0}^m \lambda_j = 1\right\}
\end{equation}
We denote $\mathbf{c}_j \in \mathcal{CH}$ the evaders that form the convex hull, i.e., $\lambda_j \neq 0$, and $\mathcal{C} = \{\mathbf{c}_j | \lambda_j \neq 0\}_0^m$. Fig.~\ref{fig:assigment}b shows the convex hull of an illustrative herd configuration.  The convex hull is important because it delimits the herd, so the smaller the area of $\mathcal{C}$, the more compact is the herd formation. It also characterizes whether an evader has escaped from the herders or not. In particular, we use the Quickhull Algorithm~\cite{barber1996quickhull} for a fast computation of $\mathcal{CH}$ and $\mathcal{C}$.

With $\mathcal{C}$, we are interested in finding a uniform partition of these evaders. A uniform partition allows to evenly cover all the herd and helps in avoiding escapes. Inspired by Voronoi diagrams~\cite{aurenhammer1991voronoi}, we propose a K-Means clustering~\cite{likas2003global} over the $\mathcal{C}$. Formally, the objective is to find the set of clusters $\mathcal{K} = \{\mathcal{K}_1, \hdots, \mathcal{K}_p | \mathcal{K}_1 \cup \hdots \cup \mathcal{K}_p = \mathcal{C}\}$ such that
\begin{equation}\label{eq:k-means}
    \mathcal{K} = \underset{\hat{\mathcal{K}}}{\arg \min} \sum_{k=0}^p \sum_{\mathbf{k}_l \in \hat{\mathcal{K}}_k} ||\mathbf{k}_l - \mu_k||^2.
\end{equation}
Here, $\mathcal{K}_k = \{\mathbf{k}_l\}_0^{p_k}$ is the set of convex-hull evaders that belong to cluster $k$, $\mu_k$ is the centroid of cluster $\mathcal{K}_k$, $p_k>0$ is the number of evaders in cluster $k$ and $p = \min(|\mathcal{C}|, n)$. The latter means that the strategy selects as many evaders as herders, unless $|\mathcal{C}| < n$, where we select all the evaders in the convex hull. Note that the K-Means algorithm~\cite{dempster1977maximum} gives the Voronoi centers $\mu_k$ $\forall k$. Fig.~\ref{fig:assigment}c shows the resulting clusters in an illustrative herd configuration. 

The final step is to select the evaders to be directly herd. The centers of the clusters can not be chosen because their dynamics are not the ones modeled in Eq.~\eqref{eq:base}. For each cluster, the furthest evader to the cluster's center is computed. This promotes the compactness of the herd. Formally, we solve
\begin{equation}\label{eq:selection}
    \mathbf{e}_k = \underset{\mathbf{k}_l \in \mathcal{K}_k}{\arg \max} \hbox{ }||\mathbf{k}_l - \mu_k||, \hbox{ } \forall k.
\end{equation}
After solving~\eqref{eq:selection}, new set of evaders $\{\mathbf{e}_1, \hdots, \mathbf{e}_p\}$ is obtained. Fig.~\ref{fig:assigment}d shows the assigned evaders in an illustrative herd configuration. These evaders are the ones that are directly controlled by the herders. In other words, the controlled state employed in Eq.~\eqref{eq:diff_eq_system_u} is $\mathbf{x}=\begin{bmatrix}\mathbf{e}_{1}^{T} &\dots  & \mathbf{e}_{p}^{T} \end{bmatrix}^{T}$.

The overall dynamic assignment strategy along with Implicit Control is summarized in Algorithm~\ref{al:algorithm}. It solves the herding problem as defined in Definition~\ref{def:control}. Roughly speaking, at each instant the algorithm ensures that the convex hull of the herd converges towards the desired reference. Therefore, since Implicit Control ensures global asymptotic stability, the convex hull converges to a region near the desired reference. In particular, the convex hull is steered to be as close as possible to $\mathbf{x}^*$. However, the repulsive forces among evaders prevents to obtain $\mathbf{e}_k = \mathbf{x}^*$ for all $k$ because there are evaders within the herd that repel them. Since at each instant the furthest evaders are selected, evaders do not escape and the center of the herd converges asymptotically to $\mathbf{x}^*$. 

\begin{algorithm}
\caption{Implicit Control and Dynamic Assignment Strategy at time $t$}\label{al:algorithm}
\begin{algorithmic}[1]
    \STATE $\mathcal{X}(t),  \mathbf{u}(t)\leftarrow$ obtain herders' and evaders' positions at instant $t$
    \STATE $\mathcal{C}(t) \leftarrow $  ConvexHull$(\mathcal{X}(t))$
    \IF{$|\mathcal{C}(t)| < n$}
        \STATE $\mathcal{K}(t) \leftarrow $ KMeans$(|\mathcal{C}(t)|, \mathcal{C}(t))$
    \ELSE
        \STATE $\mathcal{K}(t) \leftarrow $ KMeans$(n, \mathcal{C}(t))$
    \ENDIF
    \FOR{$\mathcal{K}_k(t)$ in $\mathcal{K}(t)$}
        \STATE $\mathbf{e}_k(t) \leftarrow $
        $\arg\max_{\mathbf{k}_l(t) \in \mathcal{K}_k(t)} ||\mathbf{k}_l(t) - \mu_k(t)||$
    \ENDFOR
    \STATE ${\mathbf{u}}(t) \leftarrow$ ImplicitControl$({\mathbf{e}}(t),{\mathbf{u}}(t))$
\end{algorithmic}
\end{algorithm}

It is noteworthy that we are implicitly assuming that the dynamics of the evaders are perfectly known, and that the position of all the entities can be observed. Our proposal can be directly extended to adaptive scenarios with uncertainty in the motion of the evaders~\cite{sebastian2022adaptive}, robust frameworks that recognizes the evaders~\cite{casao2021distributed} or/and a communication-aware setups by means of, e.g., a distributed Kalman filter~\cite{sebastian2021all}. 

\section{Results}\label{sec:results}

To validate the proposal, we conduct different simulated experiments. The purpose is to verify that the herding is achieved for (i) different numbers of evaders and herders, where $m \gg n$; (ii) heterogeneous groups of evaders, where each evader may have different dynamics; (iii) both fixed and time-varying references. 

The evaders can behave following two different models: the first is the one described in Eq.~\eqref{eq:PiersonBase}, called \textit{Inverse} for brevity; the second is the \textit{Exponential} that can be found in Sebastian et al.~\cite{sebastian2022adaptive}, where the only modification is that we have added the repulsion among evaders $f_j^e(\mathbf{x})$ in Eq.~\eqref{eq:PiersonBase}. The values of the parameters are: $\theta_j = 1.2$, $\gamma_j = 2 \times 10^{-4}$, $\beta_j = 0.5$, $\sigma_j=2.0$, $d_{min} = 10$m, the sample time for the simulation is $T = 40$ms, and $v_{max} = 4$m/s for both evaders and herders. The sample time is chosen for the robots to have enough time for all the calculations, while the maximum speed and the other aforementioned parameters are chosen to obtain realistic results in terms of distances, velocities and times. Nonetheless, any combination of parameters is possible. We set $\mathbf{F} = 0.25 \mathbf{I}_{2 p}$ and $\mathbf{H} = 50 \mathbf{I}_{2 p}$ in~\eqref{Eq:f_star_real} to ensure the stability of Implicit Control. We test three scenarios: (i) all evaders follow Inverse model dynamics, (ii) all evaders follow Exponential model dynamics and (iii) evaders follow either Inverse or Exponential model dynamics. When the experiments involve both models, we initialize the model of each evader randomly from a Bernoulli distribution. The evaders' positions are initialized from a normal distribution of mean $[20.0, 0.0]$m and covariance $\mathbf{I}_2$. Meanwhile, robots are deployed in an even distribution around the evaders, with center $[20.0, 0.0]$m and radius $70$m. The desired tracking reference is $\mathbf{x}^* = [-35.0,-35.0]$m and $\dot{\mathbf{x}}^* = [0.0,0.0]$m/s for $t \in [0.0, 160.0)$s and $\dot{\mathbf{x}}^* = [\nu, \omega A \cos(\omega t T)]$m/s for $t \in [160.0, 400.0]$s, with $\nu = -0.2$m/s, $A = 5$m/rad and $\omega = 0.1$rad/s.

\begin{table}[!ht]
\renewcommand{\arraystretch}{1.64}
\centering
\caption{List of symbols in the figures.}
\begin{tabular}{|c|c|}
    \hline
    Symbol
    &
    Meaning
    \\
    \hline
    \raisebox{-0.2\totalheight}{\includegraphics[width=0.08\columnwidth]{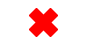}}
    &
    \raisebox{-0.2\totalheight}{{\footnotesize Desired position}}
    \\
    \hline
    \begin{tabular}{c|c}
    Symbol
    &
    Meaning
    \\
    \hline
    \raisebox{-.5\totalheight}{\includegraphics[width=0.10\columnwidth]{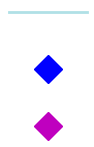}}
    &
    \begin{tabular}{c}
    {\footnotesize Herders' trajectories}
    \\
    {\footnotesize Herders' initial positions}
    \\
    {\footnotesize Herders' final positions}
    \end{tabular}
    \end{tabular}
    &
    \begin{tabular}{c|c}
    Symbol
    &
    Meaning
    \\
    \hline
    \raisebox{-.5\totalheight}{\includegraphics[width=0.10\columnwidth]{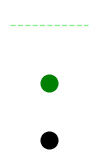}}
    &
    \begin{tabular}{c}
    {\footnotesize Evaders' trajectories}
    \\
    {\footnotesize Evaders' initial positions}
    \\
    {\footnotesize Evaders' final positions}
    \end{tabular}
    \end{tabular}
    \\
    \hline
\end{tabular}
\label{table:legend}
\end{table}

\begin{figure}[!ht]
    \centering
    \begin{tabular}{cc}
         {\footnotesize $t \in [0, 160)$s}
         &  
         {\footnotesize $t \in [160, 400]$s}
         \\
         \includegraphics[width=0.5\textwidth]{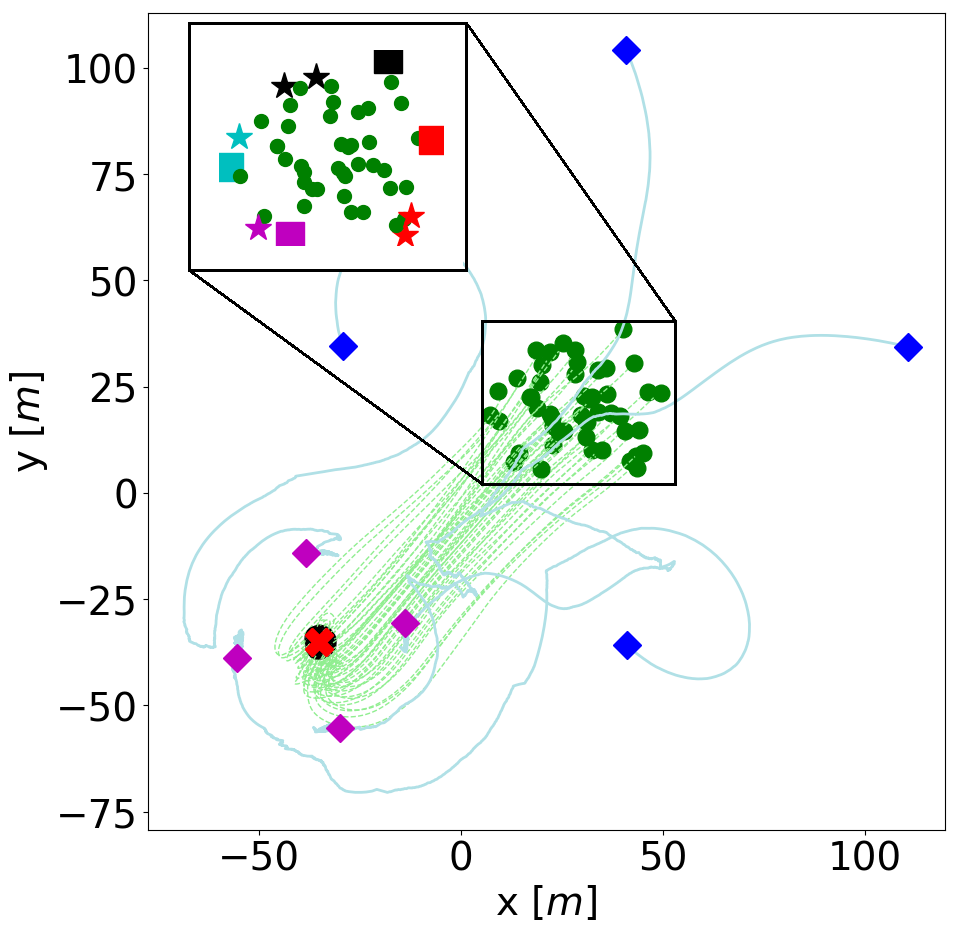}
         &
         \includegraphics[width=0.5\textwidth]{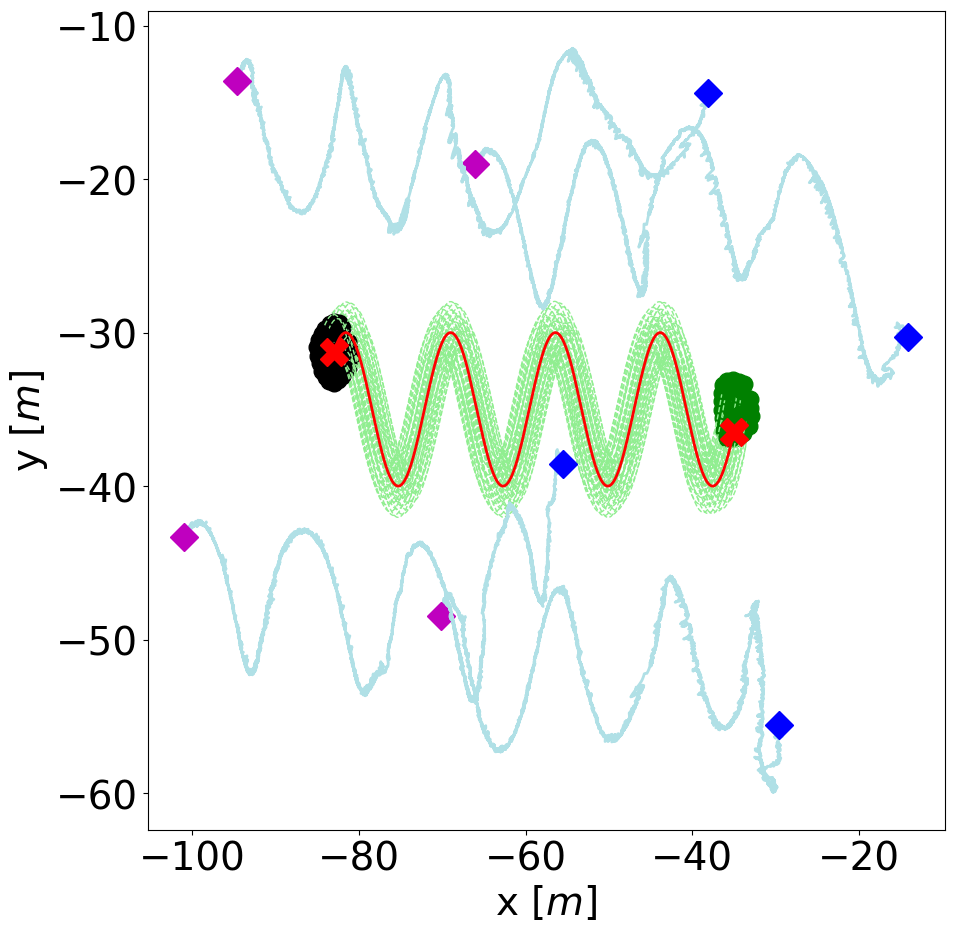}
    \end{tabular}
	\caption{Illustrative example. Herding of $50$ Inverse evaders by $4$ robotic herders. The symbols are explained in Table~\ref{table:legend}. The left pannel also shows the results of the dynamic assignment strategy for the first instant: each color represents a cluster, the square marks represent the assigned evaders while the star marks represent the other evaders that belong to the convex hull.}
	\label{fig:sim_results}
\end{figure}

\begin{figure}[!ht]
    \centering
    \begin{tabular}{cc}
         {\footnotesize Exponential evaders, $m=20$, $n=3$}
         &
         {\footnotesize Mix of evaders, $m=50$, $n=3$}
         \\
         \includegraphics[width=0.5\textwidth]{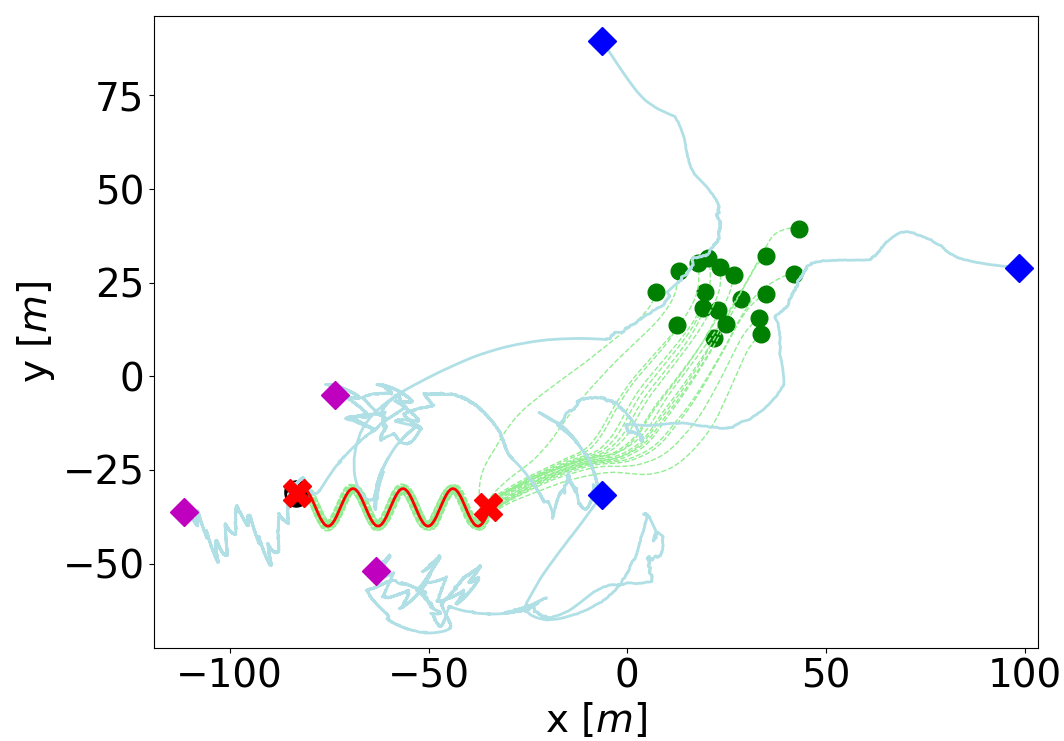}
         &
         \includegraphics[width=0.5\textwidth]{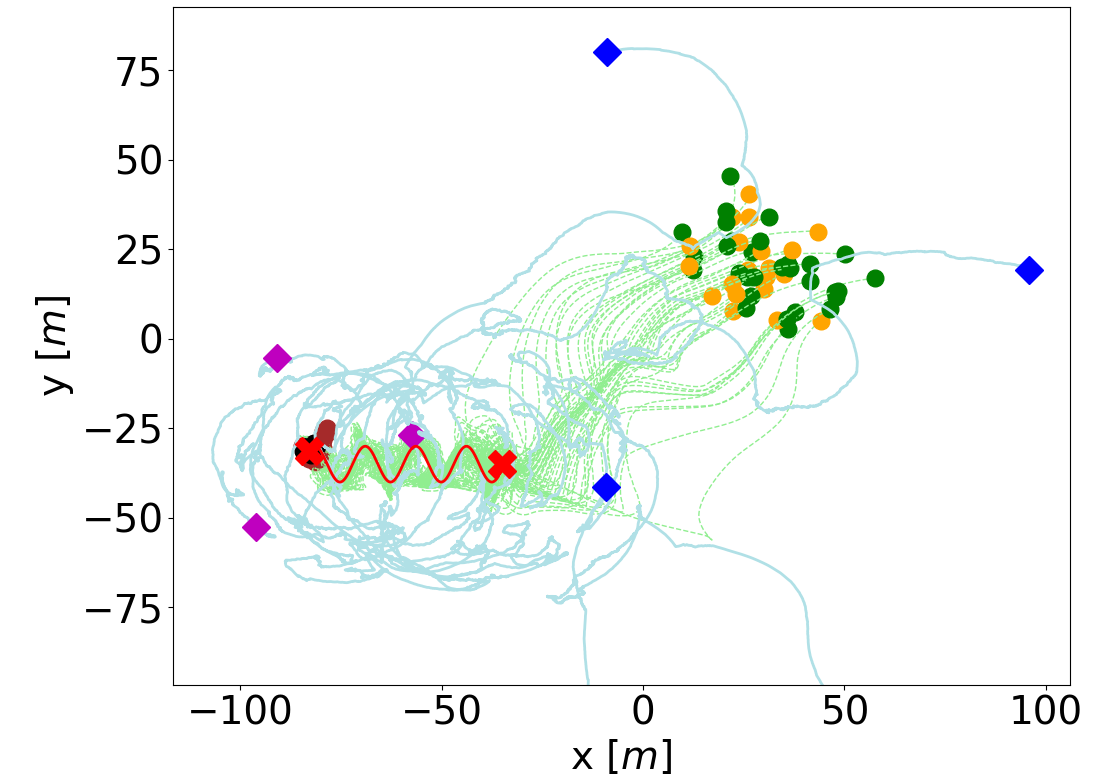}
         \\
         {\footnotesize Inverse evaders, $m=100$, $n=4$}
         &
         {\footnotesize Mix of evaders, $m=300$, $n=5$}
         \\
         \includegraphics[width=0.5\textwidth]{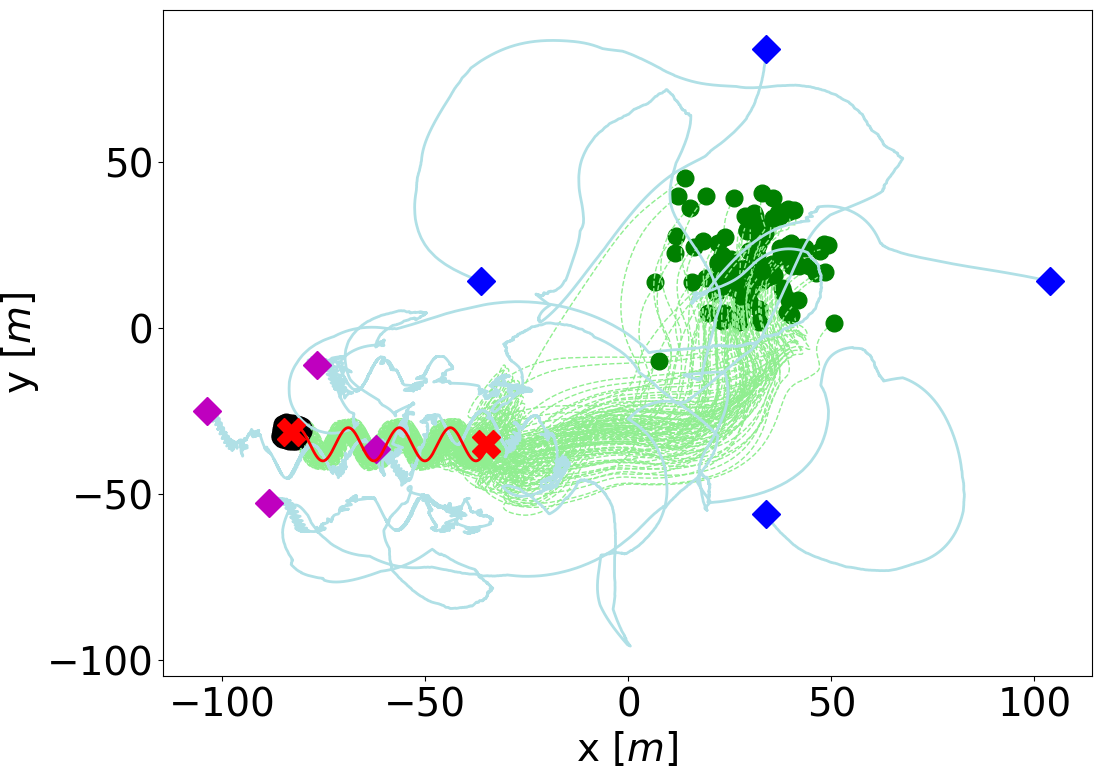}
         &
         \includegraphics[width=0.5\textwidth]{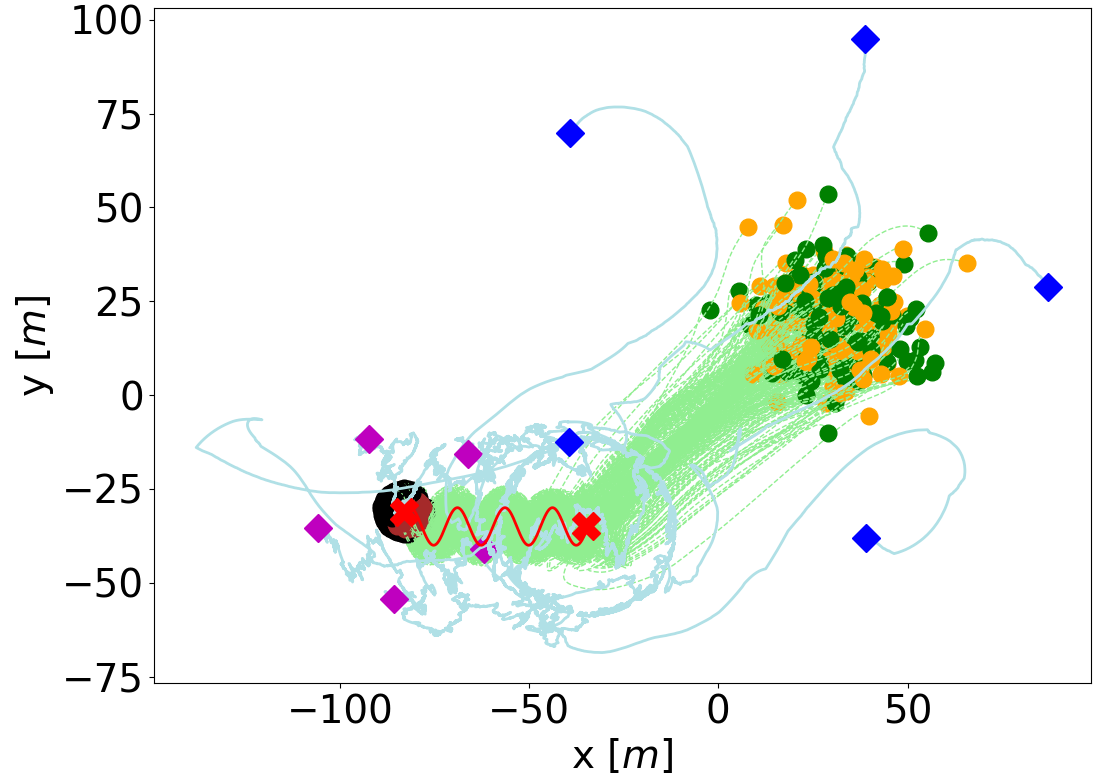}
    \end{tabular}
	\caption{Illustrative examples. The pannels depict the overall trajectories from $t=0$s to $t=400$s. In the mixed configurations, we use green and black for the initial and final configuration of the Inverse evaders whereas orange and brown for the initial and final configuration of the Exponential evaders. The other symbols follow Table~\ref{table:legend}.}
	\label{fig:sim_results2}
\end{figure}

Fig.~\ref{fig:sim_results} shows $4$ robotic herders herding $50$ Inverse model evaders. In Fig.~\ref{fig:sim_results}a it is depicted the first part of the herding, where the four herders are successful in steering all the evaders towards the desired static reference. From the trajectories it is verified that none of the evaders escape and they all remain within the influence of the herders. The convex hull is pushed towards the desired reference. Once the evaders are close to the desired location, the herders move to make the herd more compact. The final configuration of the herd is the compact black circle in Fig.~\ref{fig:sim_results}a, with the centroid in $\mathbf{x}^*$ and with the herders in a symmetric position to preserve the convex hull compactness around $\mathbf{x}^*$. After $160$s the time-varying herding begins, and in Fig.~\ref{fig:sim_results}b it is demonstrated that the herders are capable of steering the herd throughout a complex tracking reference. The compactness is preserved, and the centroid of the herd accurately follows $\mathbf{x}^*$. Therefore, the herding is achieved in the sense of Definition~\ref{def:control}. 
To complement Fig.~\ref{fig:sim_results}, Fig.~\ref{fig:sim_results2} depicts the same experiment but with either Inverse, Exponential or a mix of Inverse and Exponential evaders in different configurations. The results are as accurate as in the previous example. 

Lastly, to assess the ability for herding countless numbers of evaders, we conduct a series of simulations where the number of herders and evaders changes. We define two metrics to quantify the success of the herding:
\begin{equation}\label{eq:metrics}
\begin{aligned}
    \mathcal{L}_{\mu} =& \left|\left| \frac{1}{t_1-t_0}\int_{t_0}^{t_1} \mathbf{x}_c(s) - \mathbf{x}^*(s) ds \right|\right|,
    \\
    \mathcal{L}_{\sigma} =& \left|\left|\frac{1}{t_1-t_0}\int_{t_0}^{t_1} \frac{1}{m}\sum_{j=0}^m((\mathbf{x}_j(s)-\mathbf{x}_c(s)) - \mathbf{x}^*(s))^2 ds \right|\right|.
\end{aligned}
\end{equation}
$\mathcal{L}_{\mu}$ measures how distant is the centroid of the herd from the desired tracking reference (i.e., how far is the herd from the desired location), while $\mathcal{L}_{\sigma}$ measures how spread is the herd (i.e., if the herd remains compact or an evader has escaped). We choose $t_0=160$s and $t_1=400$s because, for a successful herding, accuracy and compactness is needed for both the steady-state static reference herding and the time-varying reference herding. The results in Table~\ref{table:results} corroborate that the solutions achieves the herding of hundreds of evaders with only a few robots. Three robots can successfully herd up to $20-50$ evaders, which is impressive taking into account that $3$ is the smallest number of herders that can cover an area ($1$ or $2$ herders form a point and a line respectively, so there is no chance of success). On the other hand, $4$ or $5$ herders can control up to $200-300$ evaders. Table~\ref{table:results} also shows that with $3$ herders and $200-300$ evaders the herd losses its coalition (large values of $\mathcal{L}_{\sigma}$) and some evaders escape from the convex hull formed by the herders, so the herd does not converge to the desired reference (large values of $\mathcal{L}_{\mu}$).  

\begin{table}[!ht]
\centering
\caption{Results with Algorithm~\ref{al:algorithm} for different numbers of evaders and herders. Green cells indicate that the herding has been successful according to Definition~\ref{def:control}; yellow cells indicate that the desired coalition stated in condition $3$ of Definition~\ref{def:control} is not achieved; and red cells indicate that the herding has not been successful according to Definition~\ref{def:control}, violating conditions $1$, $2$ and $3$.}
\begin{tabular}{|c|}
    \hline
    Inverse evaders, $\mathcal{L}_{\mu} $ $[$m$]$ $| \mathcal{L}_{\sigma}$ $[$m$^2]$
    \\
    \hline
    \begin{tabular}{c|c|c|c|c}
    \renewcommand{\arraystretch}{0.7}\backslashbox{$m$}{$n$} \renewcommand{\arraystretch}{1.5}& $3$ & $4$ & $5$ & $6$
    \\
    \hline
    $10$ &\cellcolor{LightGreen} $4.83$e${-1} | 4.18$e${-1} $ &\cellcolor{LightGreen} $1.12$e${-3} | 1.46$e${-3} $ &\cellcolor{LightGreen} $3.96$e${-4} | 1.31$e${-3} $ &\cellcolor{LightGreen} $9.82$e${-4} | 1.14$e${-3} $ 
    \\
    \hline
    $50$ &\cellcolor{LightGreen} $3.27$e${-1} | 6.88$e${-1} $ &\cellcolor{LightGreen} $3.11$e${-4} | 1.16$e${-2} $ &\cellcolor{LightGreen} $9.75$e${-4} | 1.10$e${-2} $ &\cellcolor{LightGreen} $1.29$e${-3} | 1.08$e${-2} $ 
    \\
    \hline
    $100$ &\cellcolor{LightYellow} $5.17$e${-1} | 1.52$e${-0} $ &\cellcolor{LightGreen} $7.54$e${-3} | 2.38$e${-2} $ &\cellcolor{LightGreen} $3.64$e${-4} | 2.32$e${-2} $ &\cellcolor{LightGreen} $1.10$e${-3} | 2.28$e${-2} $
    \\
    \hline
    $200$ &\cellcolor{LightRed} $1.52$e${-0} | 4.54$e${-0} $ &\cellcolor{LightYellow} $4.48$e${-1} | 4.34$e${-0} $ &\cellcolor{LightGreen} $6.57$e${-4} | 4.76$e${-2} $ &\cellcolor{LightGreen} $7.28$e${-4} | 4.71$e${-2} $
    \\
    \hline
    $300$ &\cellcolor{LightRed} $1.77$e${-0} | 7.00$e${-0} $ &\cellcolor{LightRed} $1.00$e${-0} | 3.90$e${-0} $ &\cellcolor{LightGreen} $2.25$e${-2} | 7.10$e${-2} $ &\cellcolor{LightGreen} $6.87$e${-4} | 7.23$e${-2} $
    \end{tabular}
    \\
    \hline
    Exponential evaders, $\mathcal{L}_{\mu} $ $[$m$]$ $| \mathcal{L}_{\sigma}$ $[$m$^2]$
    \\
    \hline
    \begin{tabular}{c|c|c|c|c}
    \renewcommand{\arraystretch}{0.7}\backslashbox{$m$}{$n$} \renewcommand{\arraystretch}{1.5}& $3$ & $4$ & $5$ & $6$
    \\
    \hline
    $10$ & \cellcolor{LightGreen}$4.75$e${-5} | 2.26$e${-3} $ &\cellcolor{LightGreen} $1.16$e${-3} | 1.68$e${-3} $ &\cellcolor{LightGreen} $2.94$e${-4} | 1.38$e${-3} $ &\cellcolor{LightGreen} $8.54$e${-6} | 1.35$e${-3} $
    \\
    \hline
    $50$ &\cellcolor{LightGreen} $6.43$e${-4} | 1.32$e${-2} $ &\cellcolor{LightGreen} $1.06$e${-3} | 1.11$e${-2} $ &\cellcolor{LightGreen} $1.24$e${-3} | 1.10$e${-2} $ &\cellcolor{LightGreen} $5.01$e${-5} | 1.09$e${-2} $  
    \\
    \hline
    $100$ &\cellcolor{LightRed} $4.59$e${-0} | 3.10$e${-0} $ &\cellcolor{LightRed} $2.54$e${-0} | 3.44$e${-0} $ &\cellcolor{LightGreen} $7.15$e${-5} | 2.28$e${-2} $ &\cellcolor{LightGreen} $2.18$e${-4} | 2.29$e${-2} $
    \\
    \hline
    $200$ &\cellcolor{LightRed} $3.72$e${-0} | 6.56$e${-0} $ &\cellcolor{LightGreen} $1.24$e${-3} | 1.10$e${-2} $ &\cellcolor{LightGreen} $3.47$e${-4} | 4.67$e${-2} $ &\cellcolor{LightGreen} $6.21$e${-4} | 4.68$e${-2} $ 
    \\
    \hline
    $300$ &\cellcolor{LightRed} $7.42$e${-0} | 1.15$e${+1} $ &\cellcolor{LightRed} $7.39$e${-0} | 1.02$e${+1} $ &\cellcolor{LightRed} $4.76$e${-0} | 1.19$e${+1} $ &\cellcolor{LightGreen} $2.34$e${-4} | 7.10$e${-2} $
    \end{tabular}
    \\
    \hline
    Mix of evaders,  $\mathcal{L}_{\mu} $ $[$m$]$ $| \mathcal{L}_{\sigma}$ $[$m$^2]$
    \\
    \hline
    \begin{tabular}{c|c|c|c|c}
    \renewcommand{\arraystretch}{0.7}\backslashbox{$m$}{$n$} \renewcommand{\arraystretch}{1.5}& $3$ & $4$ & $5$ & $6$
    \\
    \hline
    $10$ & \cellcolor{LightGreen} $2.47$e${-2} | 8.03$e${-3} $ & \cellcolor{LightGreen} $7.38$e${-2} | 5.28$e${-2} $ & \cellcolor{LightGreen} $6.51$e${-3} | 2.83$e${-3} $ & \cellcolor{LightGreen} $1.73$e${-3} | 2.87$e${-3} $
    \\
    \hline
    $50$ & \cellcolor{LightYellow} $1.84$e${-1} | 1.87$e${-0} $ & \cellcolor{LightGreen} $1.31$e${-1} | 7.53$e${-1} $ & \cellcolor{LightGreen} $8.57$e${-3} | 1.82$e${-2} $ & \cellcolor{LightGreen} $2.92$e${-3} | 1.52$e${-2} $ 
    \\
    \hline
    $100$ & \cellcolor{LightRed} $8.78$e${-1} | 2.39$e${-0} $ & \cellcolor{LightGreen} $3.88$e${-2} | 7.52$e${-2} $ & \cellcolor{LightGreen} $2.13$e${-2} | 3.37$e${-2} $ & \cellcolor{LightGreen} $1.46$e${-4} | 2.38$e${-2} $
    \\
    \hline
    $200$ & \cellcolor{LightRed} $2.84$e${-0} | 5.47$e${-0} $ & \cellcolor{LightGreen} $9.86$e${-3} | 8.04$e${-2} $ & \cellcolor{LightGreen} $2.40$e${-2} | 6.95$e${-2} $ & \cellcolor{LightGreen} $6.17$e${-4} | 4.82$e${-2} $ 
    \\
    \hline
    $300$ & \cellcolor{LightRed} $4.43$e${-0} | 1.40$e${+1} $ & \cellcolor{LightRed} $1.09$e${-0} | 9.81$e${-0} $ & \cellcolor{LightGreen} $7.92$e${-4} | 7.28$e${-2} $ & \cellcolor{LightGreen} $6.95$e${-3} | 9.72$e${-2} $
    \end{tabular}
    \\
    \hline
\end{tabular}
\label{table:results}
\end{table}

As a last remark, the computational time of the proposal does not vary significantly when the number of evaders increases. The only part that depends on the size of the herd is the computation of the convex hull. However, the Quickhull Algorithm~\cite{barber1996quickhull} has an average complexity $O(n \log n)$, so it can compute the convex hull of up to thousands of evaders in a few milliseconds. After that, the number of evaders involved in the computations is bounded by the number of herders, and since we only use a few herders, the proposal is cheap to compute. This is why $T=40$ms is enough for computing Algorithm~\ref{al:algorithm}.

\section{Conclusions}\label{sec:conclusions}

This paper has presented a novel control strategy to solve the herding problem of massive herds in multi-robot systems. This strategy is based on two main principles. The first one is Implicit Control, a recent general control technique for input-nonaffine systems that allows to cope with the non-cooperative, reactive and heterogeneous nature of the evaders. However, its application is limited to small herds. The second principle tackles this issue by proposing a new dynamic assignment policy. At each instant, herders choose the worthiest evaders to control in terms of coalition. This implies that, at each instant, herders directly control as many evaders as herders, preserving the conditions of Implicit Control. Besides, the strategy guarantees that the evaders remain in the convex hull and that it evolves towards the desired tracking trajectory. Therefore, all the evaders are either directly or indirectly controlled. The success of the proposal is tested in different simulations, achieving the herding of hundreds of evaders by employing just $3$ to $6$ herders.

\section*{Acknowledgments}\label{sec:acknowledgments}

This work has been supported by the ONR Global
grant N62909-19-1-2027, the Spanish projects PID2021-125514NB-I00, PID2021-124137OB-I00 and PGC2018-098719-B-I00 (MCIU / AEI / FEDER, UE), DGA T45-20R, and Spanish grant FPU19 - 05700.

\bibliographystyle{splncs_srt.bst}
\bibliography{biblio.bib}

\end{document}